# FlowsDT: A Geospatial Digital Twin for Navigating Urban Flood Dynamics

Debayan Mandal[1,2], Lei Zou[1*], Abhinav Wadhwa[4], Rohan Singh Wilkho[3], Zhenhang Cai[5], Bing Zhou[1], Xinyue Ye[5], Galen Newman[5], Nasir Gharaibeh[3], Burak Güneralp[6]

[1]GEAR Lab, Department of Geography, Texas A&M University, College Station, United States

[2]School of Geographical Sciences and Urban Planning, Arizona State University, Tempe, United States

[3]Zachry Department of Civil and Environmental Engineering, Texas A&M University, College Station, United States

[4]Discovery Partners Institute, University of Illinois System, Chicago, United States

[5]Department of Landscape Architecture and Urban Planning, Texas A&M University, College Station, United States

[6]Department of Geography, Texas A&M University, College Station, United States

*Corresponding Author: lzou@tamu.edu

## Abstract

Communities worldwide increasingly confront flood hazards intensified by climate change, urban expansion, and environmental degradation. Addressing these challenges requires real-time flood analysis, precise flood forecasting, and robust risk communications with stakeholders to implement efficient mitigation strategies. Recent advances in hydrodynamic modeling and digital twins afford new opportunities for high-resolution flood modeling and visualization at the street and basement levels. Focusing on Galveston City, a barrier island in Texas, U.S., this study created a geospatial digital twin (GDT) supported by 1D-2D coupled hydrodynamic models to strengthen urban resilience to pluvial and fluvial flooding. The objectives include: (1) developing a GDT (FlowsDT-Galveston) incorporating topography, hydrography, and infrastructure; (2) validating the twin using historical flood events and social sensing; (3) modeling hyperlocal flood conditions under 2-, 10-, 25-, 50-, and 100-year return period rainfall scenarios; and (4) identifying at-risk zones under different scenarios. This study employs the PCSWMM to create dynamic virtual replicas of urban landscapes and accurate flood modeling. By integrating LiDAR data, land cover, and storm sewer geometries, the model can simulate flood depth, extent, duration, and velocity in a 4-D environment across different historical and design storms. Results show buildings inundated over one foot increased by 5.7% from 2- to 100-year flood. Road inundations above 1 foot increased by 6.7% from 2- to 100-year floods. The proposed model can support proactive flood management and urban planning in Galveston; and inform disaster resilience efforts and guide sustainable infrastructure development. The framework can be extended to other communities facing similar challenges.

## Keywords

## 1   Introduction

Floods has accounted for over one-third of global economic losses and two-thirds of the people affected by natural hazards in recent decades (Xia et al., 2019). Recently, both the frequency and intensity of floods have been on a constant rise, leading to increasing fatalities, injuries, infrastructural damages, and economic losses worldwide (World Disasters Report, 2020). This trend has disproportionately impacted urban and peri-urban areas, which face compounded challenges from climate change, rapid population growth, and environmental degradation (Duy et al., 2018). As urban expansion continues, the need for effective strategies to reduce flood risk and enhance flood resilience becomes more pronounced (Güneralp et al., 2015).

To address natural disasters, triggered by natural events such as floods, the United Nations adopted the Sendai Framework for Disaster Risk Reduction (Macatulad & Biljecki, 2024). The Sendai Framework outlines four priorities to reduce disaster risk and enhance resilience: (i) improving disaster risk understanding through modeling interactions of different systems within human communities, (ii) investing in disaster risk reduction by building resilience through scenario planning and adaptation to various precipitation scenarios and intervention strategies, (iii) enhancing disaster preparedness and response through real-time data analysis, monitoring, and forecasting, and (iv) strengthening disaster risk governance by involving stakeholders in maintaining dynamically updated models. For urban flood resilience, these priorities underscore the need for comprehensive flood risk modeling and understanding, proactive flood resilience planning, real-time flood monitoring, response, and prediction, and robust flood governance with stakeholder engagement.

However, several challenges hinder the effective implementation of the Sendai framework for enhancing flood resilience. First, accurately delineating urban flood dynamics requires the integration of one-dimensional (1D) runoff systems (e.g., rivers and channels) with two-dimensional (2D) surface flows to capture the full scope of flood behavior. However, coupling 1D and 2D models remains technically complex. Second, high-resolution and realistic modeling is essential to reflect the intricate flood conditions in urban landscapes, but it demands substantial computational power and precise and accurate data inputs (Ye et al., 2021). Further, simulating "what-if" scenarios to anticipate the impacts of various intervention strategies and precipitation changes requires advanced modeling frameworks that can predict and visualize these conditions across space and over time. Despite a robust body of literature on urban flood resilience, few studies have directly addressed all these challenges.

Recent advances in hydrodynamic modeling and geospatial digital twin techniques present significant potential for addressing the above challenges. Enhanced hydrodynamic models support the integration of 1D and 2D flow dynamics, enabling the simultaneous representation of channelized runoff and overland flow processes. Geospatial digital twins augment these capabilities by dynamically incorporating real-time and user-defined streams, such as precipitation, water level measurements, and land surface changes—into model workflows. These systems facilitate the adaptive visualization of evolving flood conditions through immersive, spatially explicit interfaces and support scenario-based simulations to explore "what-if" contingencies.

This research develops a geospatial digital twin, integrating high-resolution geographic data, 1D-2D coupled hydrodynamic models, street-level flood simulations, and immersive visualization techniques. The Galveston City in Texas, U.S., is used as a case site because it has long faced the persistent challenge of frequent flooding. The objectives are to: (1) develop a comprehensive geospatial **FLO**od **W**atching and **S**imulation **D**igital **T**win of **Galveston** (FlowsDT-Galveston) framework that integrates topography, hydrography, and built environment; (2) validate the geospatial digital twin using historical flood events and ground-truth from web-harvested data; (3)

utilize the geospatial digital twin to simulate hyperlocal urban floods under 2yr, 10yr, 25yr, 50yr, 100yr return period rainfall scenarios; and (4) identify at-risk zones and flood impacts under various rainfall scenarios. FlowsDT-Galveston leverages a hydrologic & hydraulic (H&H) model, urban infrastructure information, and detailed topographical data to simulate flood scenarios with greater accuracy.

The novelty of this study lies in the design and implementation of a city-scale flood modeling framework with ultra-high spatiotemporal resolution, capable of supporting real-time simulation through potential integration with sensor-based data collection systems. The resulting geospatial digital twin provides a dynamic computational environment for simulating and visualizing street-level flood conditions, enabling the identification of high-risk assets and potential property damages. Beyond its analytical capabilities, the digital twin functions as an interactive platform that supports multi-stakeholder engagement. It facilitates collaboration among urban planners, emergency management agencies, policymakers, and residents, promoting the co-design and co-implementation of resilience-oriented strategies and preparedness measures. In addition to the ultra-high spatiotemporal resolution and real-time simulation capabilities, this study advances the Level of Detail (LoD) in urban digital twins. While most existing Urban Digital Twins (UDTs) operate at LoD3—capturing above-ground infrastructure—our integration of Galveston's detailed underground storm sewer network (including inlets, outlets, conduits, and junctions) effectively elevates the model to LoD4, significantly enhancing its functional resolution and utility in urban flood analysis (Deng et al., 2021; Liu et al., 2023). Furthermore, by explicitly modeling flow pathways and obstructions at the street scale, the framework provides actionable insights for identifying non-point source (NPS) pollutant transmission during flood events—an increasingly critical concern for public health and water quality. Such detailed hydrodynamic modeling offers new opportunities for targeted flood mitigation interventions and infrastructure planning (Hou et al., 2021; Moore et al., 2017). It facilitates collaboration among urban planners, emergency management agencies, policymakers, and residents, promoting the co-design and co-implementation of resilience-oriented strategies and preparedness measures. This integrative and participatory approach enhances both the scientific robustness and the practical utility of urban flood risk management.

The structure of this article is as follows. Section 2 describes the conceptualization and historical context of digital twins, interpreting their significance within the field of urban stormwater management. Section 3 elaborates on the selected study area of Galveston, detailing its specific environmental conditions, the effectiveness of the current drainage system, and the rationale behind its selection for the digital twin initiative. Section 4 catalogs the data sets and methodology employed to develop FlowsDT-Galveston. Section 5 presents the outcomes and applications of the digital twin based on historical events, designated rainfall scenarios, and real-time flood forecasting through radar data. Finally, Section 6 concludes the implications of this work and sheds light on future directions.

## 2  Background

### 2.1  Hydrodynamic Modeling

Previous research to enhance urban flood prediction and resilience with focus on the development of numerical models to characterize hydrodynamic behavior and simulate inundation patterns. Two dominant methodological paradigms have emerged: data-science and physics-based models. Data-science approaches employ Machine Learning (ML) algorithms to model floods based on weather conditions and the built-environment without considering the constraints of physical processes (Liang et al., 2023). For instance, ML models trained on historical datasets have been used to estimate flood frequencies and assess risk under various climatic scenarios (Engeland et al., 2018). While these models can offer rapid assessments and reduced computational overhead, their performance is often constrained by the availability, resolution, and reliability of historical and observational data (Saber et al., 2023).

Physics-based models use physics principles and equations to describe hydrodynamics. They can be applied to a wide range of environmental conditions and scales with minimal calibration (Camporese & Girotto, 2022). Physics-based hydrodynamics models are further differentiated as hydrological and hydraulic models. Hydrological models simulate the rainfall-runoff processes while hydraulic models use runoff to simulate the movement of water through rivers and floodplains. Early hydrological models, such as the Rational Method and Unit Hydrograph Method, use simple empirical formulas to estimate peak discharge from rainfall events but lack accuracy in complex watersheds with varied topography and land use (Chow et al., 1988). Traditional hydraulic models, including the Hydrologic Engineering Center's River Analysis

System (HEC-RAS) (Costabile et al., 2020), were primarily one-dimensional (1D) simulating flow along a single channel dimension. These traditional models struggle to accurately represent floodplain inundation and urban flooding where multi-directional flows occur (X. Wu et al., 2018).

Recently, more sophisticated physics-based flood models were developed, allowing for a more detailed representation of floodplain dynamics and urban flooding (McMillan & Brasington, 2007; Zhao et al., 2021). Hydrological models, such as the Soil Conservation Service (SCS) Curve Number method, were designed to consider various hydrological processes, such as infiltration, evapotranspiration, and surface runoff. Meanwhile, hydraulic models have evolved from 1D finite element method (FEM) routing to two-dimensional (2D) finite volume method (FVM) routing to simulate water flow across grids (Schmitt & Scheid, 2020). Leveraging the strengths of both 1D and 2D hydraulic models, Yu et al., (2019) developed coupled 1D/2D distributed and physically based models. They used 1D simulations for river channels and 2D simulations for floodplains, striking a balance between computational efficiency and accuracy. These applications of these models, however, were limited to large geographical areas and relied on low-resolution datasets. This is done because large-scale high-resolution simulations have higher computational constraints (Saksena et al., 2020). Furthermore, these products are typically applied in a static manner – run under fixed input conditions and do not address the need for a proactive and real-time model under rapidly changing climatic patterns. Yet, despite these problems, the spatial patterns of flood impacts predicted by these models have been used to design urban infrastructure and formulate regulatory policies.

## 2.2   Geospatial Digital Twins

Initially developed within the aerospace and manufacturing sectors (Grieves, 2016), digital twin technologies have more recently been extended to urban systems and spatial planning contexts. The National Academies of Sciences, Engineering, and Medicine (NASEM) offers a comprehensive definition, stating that a digital twin refers to a collection of virtual information constructs that replicate the structure, context, and behavior of a natural, engineered, or social system (or system-of-systems); are continuously updated with data from their physical counterpart; possess predictive capabilities; and support decision-making processes to achieve value (Committee on Foundational Research Gaps and Future Directions for Digital Twins et al., 2024). Central to their utility is the ability to analyze and monitor systems in real-time – such as environmental changes and natural resource management, thereby facilitating complex participatory decision-making. Digital twins have been implemented across various fields, including but not limited to manufacturing (Tao et al., 2019), healthcare (C. Wu et al., 2022), smart cities (Mylonas et al., 2021), agriculture (Verdouw et al., 2021), and energy (Lamagna et al., 2021).

Geospatial digital twins represent advanced digital replicas of physical environments that integrate precise and accurate geospatial data, typically acquired through advanced data collection technologies such as aerial sensors, Light Detection and Ranging (LiDAR), and photogrammetry, to construct comprehensive virtual models (Ye et al., 2023). These models are designed to simulate and reflect the real-world characteristics and dynamics of geographic areas, providing a detailed and accurate digital counterpart to physical spaces. Geospatial digital twins enable innovative applications across diverse domains, particularly in urban settings, where they support smart city initiatives, optimize urban planning processes, and improve the efficiency of public services (Ali et al., 2023).

Geospatial digital twins for urban areas offer a promising solution to model and manage floods at hyperlocal scales. A geospatial digital twin can be integrated with different rainfall scenarios which aid in producing inundation maps to demonstrate street and basement flooding. It can incorporate temporal variations in its predictions at small time steps, making effective 4-Dimensional (4D; length, breadth, height, and time) forecasts. Additionally, employing geospatial digital twins at such fine spatiotemporal scales can simulate cascading effects under varying hazard and built-environment scenarios, allowing for 'what-if' analyses to simulate potential hazard events. Geospatial digital twins improve science communication by moving away from the traditional top-down approach, where stakeholders were only passive recipients of flood maps, to a more interactive and collaborative process. Moreover, the 4D representation of flood inundation available in geospatial digital twins engages with stakeholders at an immersive level. Such realistic visualization, which can simulate dramatic scenarios while adhering to accurate scientific representation (Kuser Olsen et al., 2018), has proven to improve the interpretation of risk and subsequent decision-making processes.

Geospatial digital twins have been successfully employed in urban planning and management. Virtual Singapore creates a dynamic 3-Dimensional (3D) model of the city, integrating real-time environmental data to enhance urban planning, infrastructure management, and public service deployment (e.g., flood management and flight safety) (Ignatius et al., 2019). The Herrenberg initiative in Germany uses geospatial digital twins with real-time data from sensors and IoT devices to improve traffic flow and urban infrastructure management (Dembski et al., 2020). The Helsinki 3D+ project provides detailed 3D models for enhancing public services and infrastructure development (City of Helsinki, 2023). Additionally, the Smart Dubai initiative (Alnaqbi & Alami, 2023) leverages geospatial digital twins to improve public safety, transportation, and sustainability in urban environments. These projects demonstrate how comprehensive data integration in geospatial digital twins enriches decision-making processes, leading to more informed and effective urban management strategies.

The evolution of digital twins and geospatial digital twins consists of three stages (Jones et al. 2020): (1) the Digital Twin Instance (DTI) - digital replicas based on one instance or event; (2) the Digital Twin Aggregate (DTA) – digital replicas that integrate multiple DTIs; and (3) the Digital Twin Environment (DTE) – digital replicas that combine diverse domain aggregates. This study focuses on capturing the essence of a Digital Twin Aggregate that encompasses not only the predictive capability of capturing the effect of designed storms but also can be plugged into real-time radar data for real-time flood forecasting. It begins with data acquisition, where diverse sources, such as sensors, satellite imagery, and databases, provide real-time and historical information about physical entities. Next, data integration and processing enable seamless interoperability across various systems. Following integration, the modeling and simulation processes allow the system to replicate real-world conditions by incorporating physics-based, statistical, or machine-learning models. These models support predictive analytics, enabling the assessment of different scenarios and system behaviors under various conditions. Subsequently, real-time data assimilation ensures that the digital twin remains dynamically updated. The next step establishes visualization and interaction mechanisms, e.g., interactive immersive environments to enhance user engagement. Finally, decision makers can leverage the aggregated data and simulations to facilitate informed, optimized decision-making.

## 3   Study Area

Galveston Texas, the study area, is an island frequently assailed by floods. The site serves as a classic example of urban areas facing significant flood risks due to its geographic and climatic conditions. Spanning a geographical expanse of 209.3 mi², the city is positioned between latitudes

29.15°N and 29.38°N, and longitudes 94.79°W and 95.02°W (Figure 1). This city, characterized by its coastal setting, is susceptible to flooding from a variety of sources, including storm surges, heavy rains, and rising sea levels.

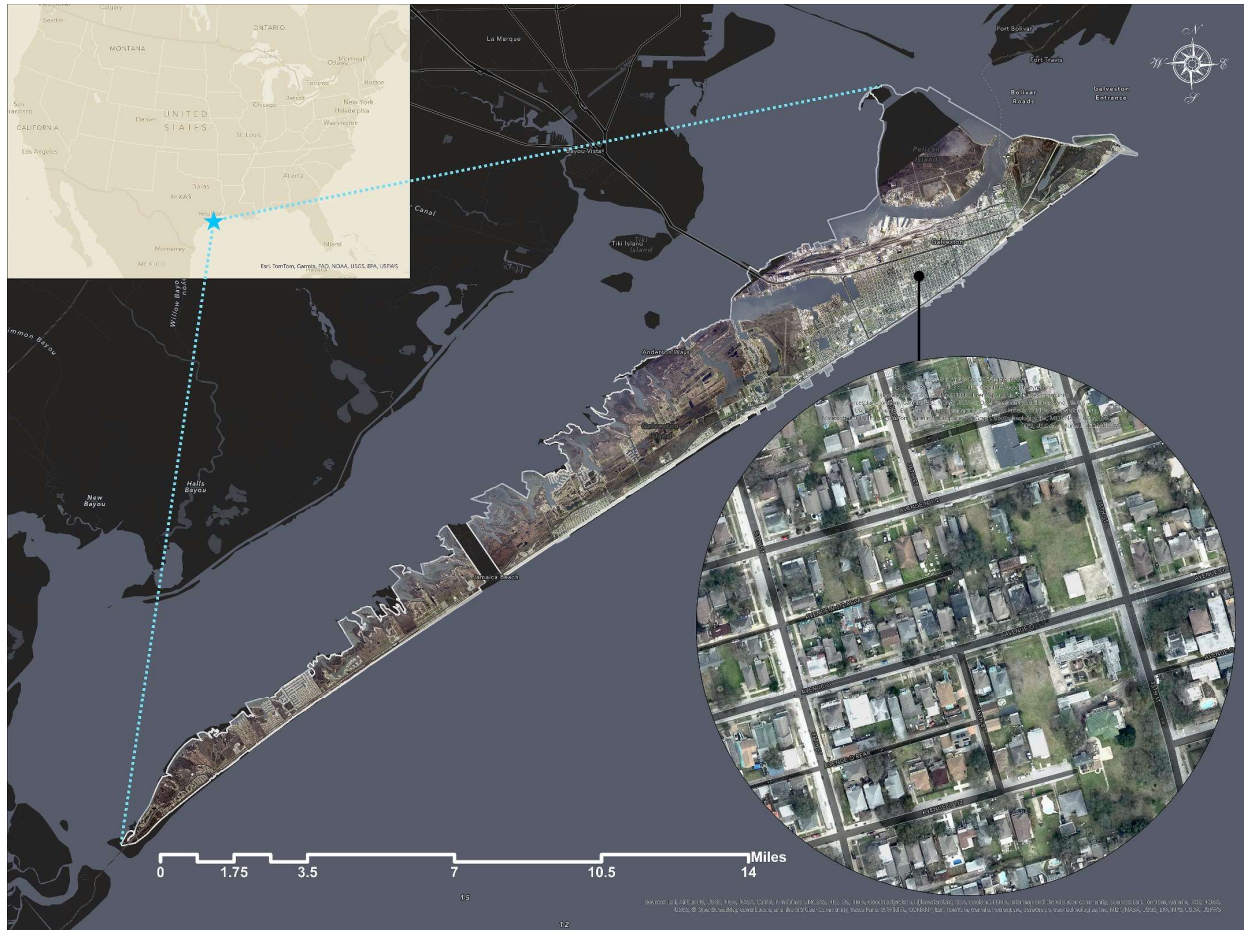

*Figure 1: The study area – the Galveston City, Texas, U.S.*

In Galveston, a huge corpus of research focuses on marine flooding (coastal flooding events driven by storm surges and elevated sea levels) (Cai et al., 2023), at the expense of pluvial and fluvial flooding, even though the latter has significantly exacerbated flood impacts alongside marine flooding during hurricanes in the city (Huang et al., 2021). For instance, if a 100-year flood (1% chance) event occurs today, it is projected to affect 26,651 properties in Galveston (First Street Foundation, 2024). In a time where the majority of homeowners opt for 30-year mortgages, an event of this magnitude has a 26% chance of occurring during the mortgage (Texas Flood Insurance, 2024). Additionally, potentially higher sea levels, erratic weather patterns, and stronger storms due to a changing climate will likely result in endangering even a larger percentage of properties. Due to these factors and the increasing urbanization (due to growing tourism), static flood risk maps may be conservative in predicting damage. The present study innovates by proposing the integration of a geospatial digital twin to accurately simulate flood risk under changing climate and urbanization scenarios.

## 4    Data and Methods

### 4.1    Overall Framework

This research leverages LiDAR, land cover characteristics, Storm Sewer dimensions, and Rainfall Gauge data to construct the geospatial digital twin model – FlowsDT-Galveston. Figure 2 shows the comprehensive workflow of this study which uses two phases: digital twin creation and digital twin utility. The digital twin creation phase consists of five steps. The first step is subcatchment delineation using the derived elevation and land use data. In urban hydrology, the urban landscape is discretized as subcatchments - small, manageable hydrological units that can capture the aspects of the local topography, land use, and soil characteristics. This contributes to the delineation of natural channels. In the second step, the storm sewer geometries are digitized and joined with natural channels to form the 1D hydraulic model. Third, the 2D mesh is generated to represent overland flow paths, with cells adapted around the built-in obstructions (buildings and infrastructure).

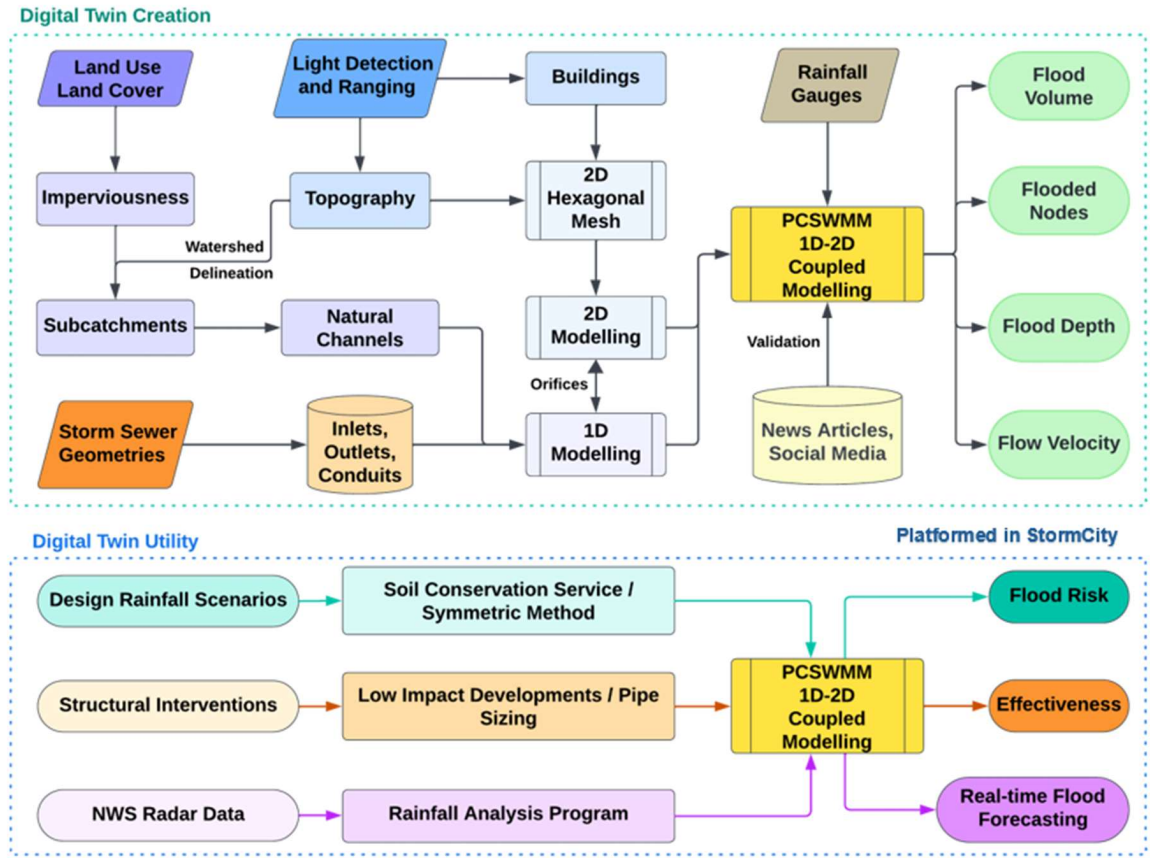

*Figure 2: The comprehensive workflow of creating and using the Digital Twin system (FlowsDT-Galveston)*

In the fourth step, the 1D and 2D models are merged using orifices to create a comprehensive hydrologic and hydraulic model, leveraging the Personal Computer Storm Water Management Model (PCSWMM). PCSWMM is a physically distributed hydrologic-hydraulic model, which can simulate runoff quantity from urban zones (Rossman, 2017; Rossman & Huber, 2015). PCSWMM allows for 1D computation in the case of channels and conduits where primarily 1D flow is observed. Additionally, it can compute flows in 2D for overland flow through a mesh and connect this model to the 1D system of the area. These can be simulated based on user-provided time-steps. Precipitation received in each subcatchment is used by the model to generate the imperviousness runoff quantity. The hydraulic functionalities of this model compute and transport the runoff through a system of conduits/natural channels, inlets, outlets, storage areas, etc. This system can dynamically simulate both the surface and sewer flow interactions, capturing the complexity of the urban landscape.

Finally, the rainfall observations through the two rain gauges (Scholes and Northern Galveston) serve as inputs to the model to obtain simulated flooding data. Flood conditions were simulated using the developed digital twin model with historical events and validated the results against ground-truth inundated areas identified through web-harvested social sensing data, i.e., social and news media reports. The complete digital twin is then housed in an immersive 3D platform created with the help of StormCity. StormCity, developed by Computational Hydraulics International[1] (CHI), is a 4D visualization platform that enables realistic, interactive representation of urban flood dynamics and infrastructure systems. It facilitates the utilization of the digital twin and allows the stakeholders to meaningfully engage with the results. The digital twin utility phase can leverage the validated model to simulate impacts of various design rainfall scenarios – hypothetical rainfall events with defined return periods like, 2-year, 10-year and 100-year storms (through predictive precipitation models, e.g., the soil conservation service method), structural interventions (like low-impact developments or pipe sizing), and real-time flood forecasting (utilizing radar data to dynamically obtain flood inundation maps). The subsequent subsections describe each step in detail.

## 4.2 Subcatchment Characterization

The 1-meter resolution LiDAR data, obtained in 2018 for Galveston City was sourced from the Texas Natural Resources Information System (TNRIS)[2]. It contains a variety of point cloud classifications, including Ground; Low, Medium, and High Vegetation; Building; Low noise; Water; Rail; Wires; and Bridge Deck. This diverse range of classification offers a comprehensive view of the landscape, allowing us to gain insights into the area's terrain, vegetation, and infrastructure. The LiDAR data was used to create a Digital Surface Model (DSM) that includes all classes (Figure 3a) and a Digital Terrain Model (DTM) that includes only ground classifications. The Normalized Digital Surface Model (nDSM) was generated by subtracting the

---

[1] https://www.chiwater.com/Home

[2] https://tnris.org/

DTM from the DSM. ArcGIS Pro was used to extract and create objects of different features such as 3D models of buildings, trees, roads, and 2D building footprints from the LiDAR data.

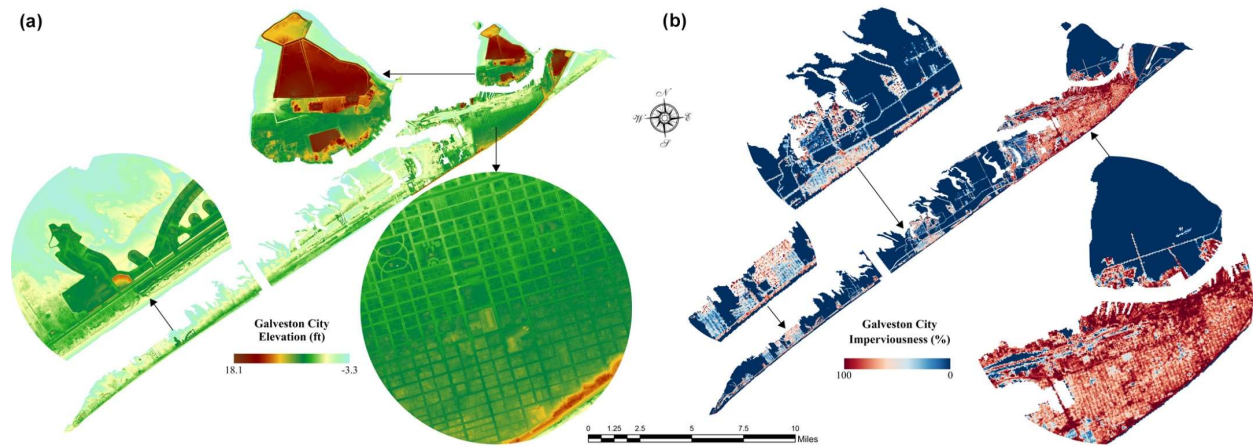

*Figure 3 (a) Digital Surface Model and (b) Imperviousness percentage of Galveston City*

The next step discretizes Galveston City into multiple subcatchments – a hydrologically distinct area draining to a common outlet – to characterize the spatial variability in overland drainage pathways, surface properties, and connections into drainage pipes and channels. The Watershed Delineation Tool in PCSWMM was used to discretize Galveston with the target level of 10 acres (40,000 square meters) to maintain the granularity while avoiding extensive computational time. This tool calculates directions and contributing areas, generates natural conduits, and creates subcatchments based on the topographical features of the DSM. The natural conveyance system is characterized using nodes (junctions) and links (conduits). In the context of PCSWMM, nodes are points where two links are joined (junctions) or where external flow enters or exits the system (inlets, outlets). The links (conduits) are connections that transport flow from one node to the other. A total of 1823 junctions, 1609 conduits and 2396 subcatchments were generated. The junctions physically denote the convergence of natural channels or flow lines dictated by the local topography.

Each discretized subcatchment consists of pervious and impervious areas with and without depression storage (i.e., the initial volume of water retained on the surface before runoff begins; by default, for pervious, it is 0.15 mm and, for impervious, 0.1 mm). We obtained impervious land area percentage obtained in 2021 from the National Land Cover Database (NLCD) which provides nation-wide impervious rates at 30-meter resolution (Figure 3b). This helps capture the subcatchment characteristics for the study area, delineating surfaces that inhibit infiltration and generate runoff. By integrating this dataset, we were able to derive precise metrics of impervious surfaces such as roads, rooftops, and other sealed structures. The analytical process involved area-averaging the imperviousness values from the NLCD to obtain the percentage of impervious and pervious areas within each subcatchment.

## 4.3    Storm Sewer Geometry

In the urban context, subsurface drainage network (Figure 4) plays a significant role in conveying runoff. Geometric characteristics of the storm sewer system, including shape, size, and connectivity, were integrated into the digital twin model to ensure a realistic representation of the urban drainage network. Data describing storm sewer systems were procured from the Galveston City's GIS data repository[3] which was created in 2004, and has been updated till 2020. This data set comprised an extensive network of 992 inlets, 1504 storm sewer mains, 59 outfalls, and 151 junctions. Inlets refer to entry points where surface water enters the storm sewer system, typically through grates or catch basins. Storm sewer mains are the primary conduits that transport stormwater away from urban areas. Outfalls are the discharge points where stormwater exits the sewer system and is released into natural water bodies like rivers, lakes, or oceans. Junctions are the connecting points within the sewer system where multiple sewer mains converge, often involving changes in direction or elevation. These inputs provided a foundational framework for constructing accurate digital representations of the urban drainage infrastructure within the PCSWMM.

---

[3] https://gis-galveston.hub.arcgis.com/

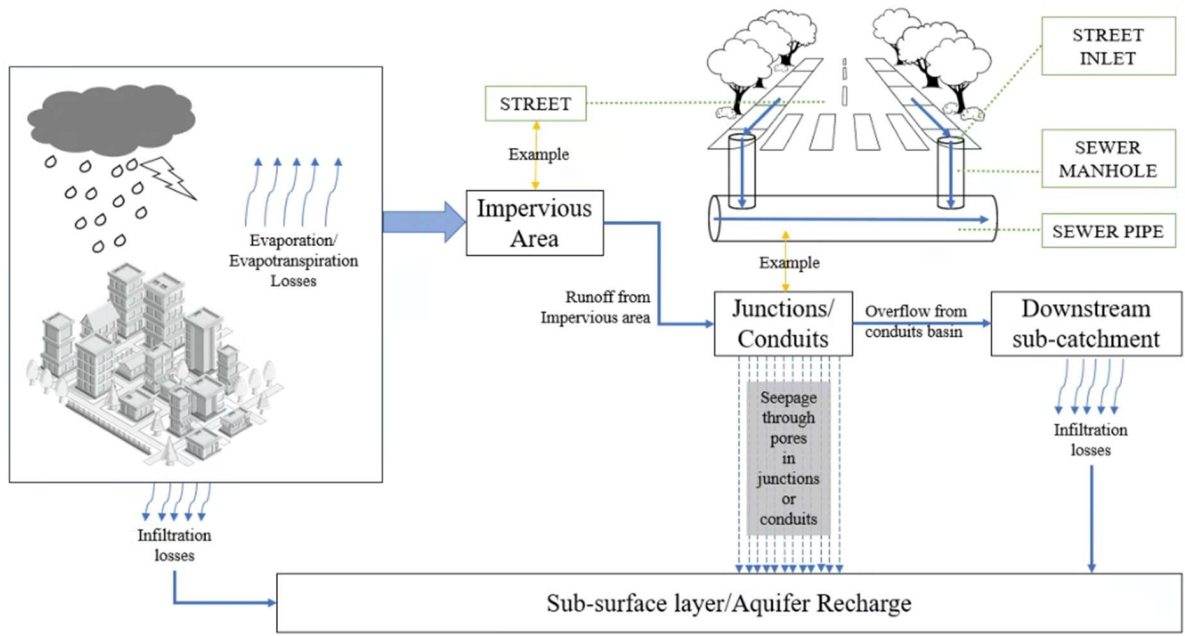

*Figure 4: Diagrammatic Representation of Urban Drainage System*

The initial dataset required rigorous correction (such as changing invert elevations, connecting missing conduits, and providing dimensions to inlets) to rectify several inconsistencies, ensuring the integrity of the model. Discrepancies such as orphaned network segments with no terminus at water bodies were excised while missing inlets were supplemented at adjoining pipe diameters. Superfluous bends in the piping that contributed unnecessary complexity were streamlined, while junctions associated with pipes having slope data of zero were standardized to maintain uniformity in depth. In cases where slope and length parameters were available, downstream junction depths were deduced from their upstream counterparts. Notably, multiple anomalous inlets, outfalls, and pipes were scrutinized and eliminated. Pipes with zero size were removed. Inlets with missing depths were given the same diameter as the adjoined conduit if it was at an orifice. Moreover, to ensure hydraulic connectivity, strategic conduit additions were implemented for orphaned networks. At certain missing datapoints, the recorded upstream and downstream crown heights of the pipes were used for establishing junction depths. Lastly, short, parallel pipes were consolidated into barrel structures with analogous dimensions, thereby optimizing the model's structural coherence. The PCSWMM's Watershed Delineation Tool (WDT) generated natural flow paths that were connected to the storm sewer junctions based on spatial proximity to make up the comprehensive 1D network of the city. Figure 5 illustrates a snapshot of the resultant network comprising 1638 conduits, 1529 junctions, 4 dividers, and 100 outfalls.

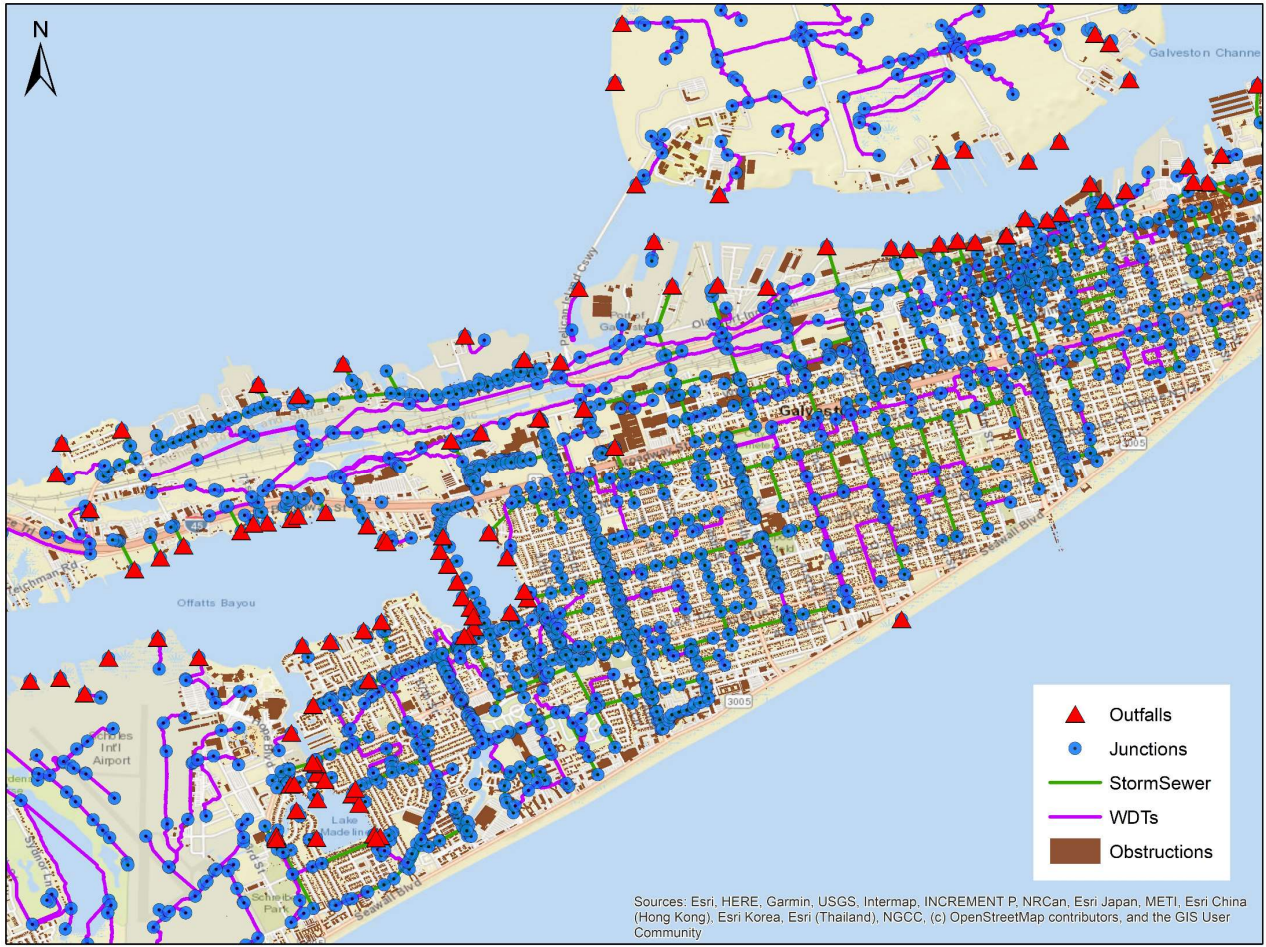

*Figure 5: A Snap of the Storm Sewer and natural drainage system of the northern portion of Galveston City*

## 4.4 2D Overland Modelling

The 2D floodplain model was used to compute overland flow once the nodes of the 1D network got overflowed due to runoff going over capacity of the conduits. This model computes flow in two dimensions over the surface through a mesh overlaying the built area. In this study, the mesh consisted of a uniformly sized hexagonal grid, with each cell area amounting to 8390 ft$^2$ (779.46 m$^2$) and a height of 30 ft (9.1 m). The buildings and structures extracted from the LiDAR data acted as obstructions to the flow. 2D nodes were first created to be the guiding points for the mesh creation. Subsequently, the mesh was laid adaptively around the obstructions. As the flood level never went above 30 ft, the default specifications worked well for the study area. Additionally, outfall nodes to the 2D mesh were created along the boundary of the island and inland lakes to provide outlets from the mesh to the lakes and ocean. A total of 6577 outfalls to the 2D mesh were created along the Galveston City boundary.

## 4.5    1D-2D Coupled Modelling and Simulation

The 1D conduit model and 2D floodplain model were connected through orifices for surface flow generation and hydrodynamic calculations of the conduit network. Physically, these orifices denote the sewer openings in the street and account for the energy loss as the flow enters the 1D system. The rainfall gauge data were then used as an input of water to the created 1D-2D network. The National Oceanic and Atmospheric Administration (NOAA) Integrated Surface Database[4] (ISD) was used to collect rain gauge data during past flood events. ISD is a global repository of multiple weather parameters such as wind speed and direction, wind gust, temperature, dew point, cloud data, sea level pressure, among others. It consists of hourly observations compiled from multiple sources. For this study, we gathered the rainfall measurements of a 24-hour period on the same day as the flood event (as per the NOAA Storm Events Database). These were converted to rainfall intensity for modelling purposes. It was assigned to the subcatchments based on the Thiessen polygon method. Our model translates rainfall events into runoff using the Manning formula (Eqn. 1), routing through the storm sewer network to identify critical flood volumes and the spatial distribution of flooded nodes.

$$Q = \left(\frac{1.49}{n}\right) A R^{\frac{2}{3}} \sqrt{S} \qquad (1)$$

where Q is the flow rate (ft$^3$/s); n is the manning roughness coefficient; A is the catchment area (ft$^2$); S is the slope of catchment area (ft/ft); and R is the hydraulic radius (ft). These parameters are determined by land use and delineated subcatchment area.

The rainfall-runoff processes and the infiltration behavior of urban catchments were simulated through the Horton infiltration model (Chow et al., 1988) (Eqn. 2):

---

[4] https://www.ncei.noaa.gov/products/land-based-station/integrated-surface-database

$$f_p = f_\infty + (f_0 - f_\infty)e^{-k_d t} \tag{2}$$

where $f_p$ is infiltration capacity into soil (ft/sec); $f_\infty$ is the minimum or equilibrium value of $f_p$ (at $t = \infty$) (ft/sec); $f_0$ = maximum or initial (pre-storm) value of $f_p$ (at $t = 0$) (ft/sec); $t$ is the time from beginning of storm (sec); and $k_d$ is the infiltration capacity decay coefficient (sec$^{-1}$). These parameters are determined using imperviousness and curve number calculations. The curve number (CN), derived from hydrologic soil groups and land use classifications, helps estimate empirically standardized infiltration capacity.

Concurrently, the dynamic wave method was employed for hydraulic modeling to capture the complex interactions of flow and storage, offering a comprehensive portrayal of the hydraulic conditions (Chow et al., 1988). The governing equations for dynamic wave routing used for hydrodynamic actuation are the Saint Venant Equations (Eqns. 3-4).

Continuity (Conservation of Mass):

$$\frac{\partial A}{\partial t} + \frac{\partial Q}{\partial x} = 0 \tag{3}$$

Momentum (Conservation of Momentum):

$$\frac{\partial Q}{\partial t} + \frac{\partial (\frac{Q^2}{A})}{\partial x} + gA\frac{\partial H}{\partial x} + gAS_f = 0 \tag{4}$$

where $x$ is distance (ft); $t$ is time (sec); $A$ is flow cross-sectional area (ft$^2$); $Q$ is flow rate (cfs); $H$ is hydraulic head of water in the conduit (ft); $S_f$ is friction slope (head loss per unit length); and $g$ is acceleration of gravity (ft/sec$^2$). These equations describe how water moves in open channels and pipes by considering conservation of mass and momentum. Therefore, they account for how water flows, accelerates, and changes depth based on gravity, friction, and pressure. These equations help model floodwaters, river flows, and urban drainage systems by predicting how water levels and velocities evolve over time.

The model simulates water flow across the city's terrain and infrastructure, predicting flood volumes, depths, and extents per routing time step.

## 4.6 Model Validation with Social Sensing

Validation of FlowsDT-Galveston is done by comparing its outputs based on historical flood events with reported flooded areas from social and news media during those events. The NOAA Storm Events (SE) Database (https://www.ncdc.noaa.gov/stormevents/) serves as a disaster repository, chronicling weather-related phenomena across the United States, including flood incidents. The database archives a spectrum of storm events dating back to January 1950, continuously curated and updated by the National Weather Service, with a publication delay of 90 to 120 days. Alongside, SHELDUS database, a commercial hazard database hosted in Arizona State University catalogues every hazard event without aggregation per county, inclusive of flood events. Leveraging these databases, we filtered records pertinent to flood events within Galveston, Texas, spanning a decade from 2010 to 2019. We have identified and modeled four flood events on the dates – 05/14/2010, 03/06/2016, 18/04/2017 and 09/29/2018, as there were extensive data found on them and the flooding was due to heavy rainfall only (pluvial and fluvial). The cumulative rainfall during these events were 10.35 (0.27), 8.67 (0.22), 7.3 (0.19), and 8.39 (0.21) inches (meters), respectively.

To validate the 1D-2D coupled hydrologic and hydraulic model, we used the extracted location-based information from various crowdsourced datasets (Figure 6), overlaid them over the simulated flood coverages, and evaluated the prediction accuracy. Specifically, we used an integrated information harvesting system (Wilkho et al., 2023, 2024; Wilkho & Gharaibeh, 2025) to collect ground-truth information on flooded areas for past events. For a past event of interest, the system retrieves documents (social media posts, news reports, government records, among others) from the web using Google Search Application Programming Interface (API) and Bidirectional Encoder Representations from Transformers (BERT), a pre-trained large language model. It then classifies the paragraphs contained in the retrieved webpages into pertinent categories (e.g., damage, fatality, among others) and extracts specific information (e.g., impacted neighborhoods and streets) from the paragraphs classified in the damage category using fine-tuned Named Entity Recognition (NER) models. Finally, these data were used to validate the model by corresponding to these zones and the maximum flood depth level in the respective model results.

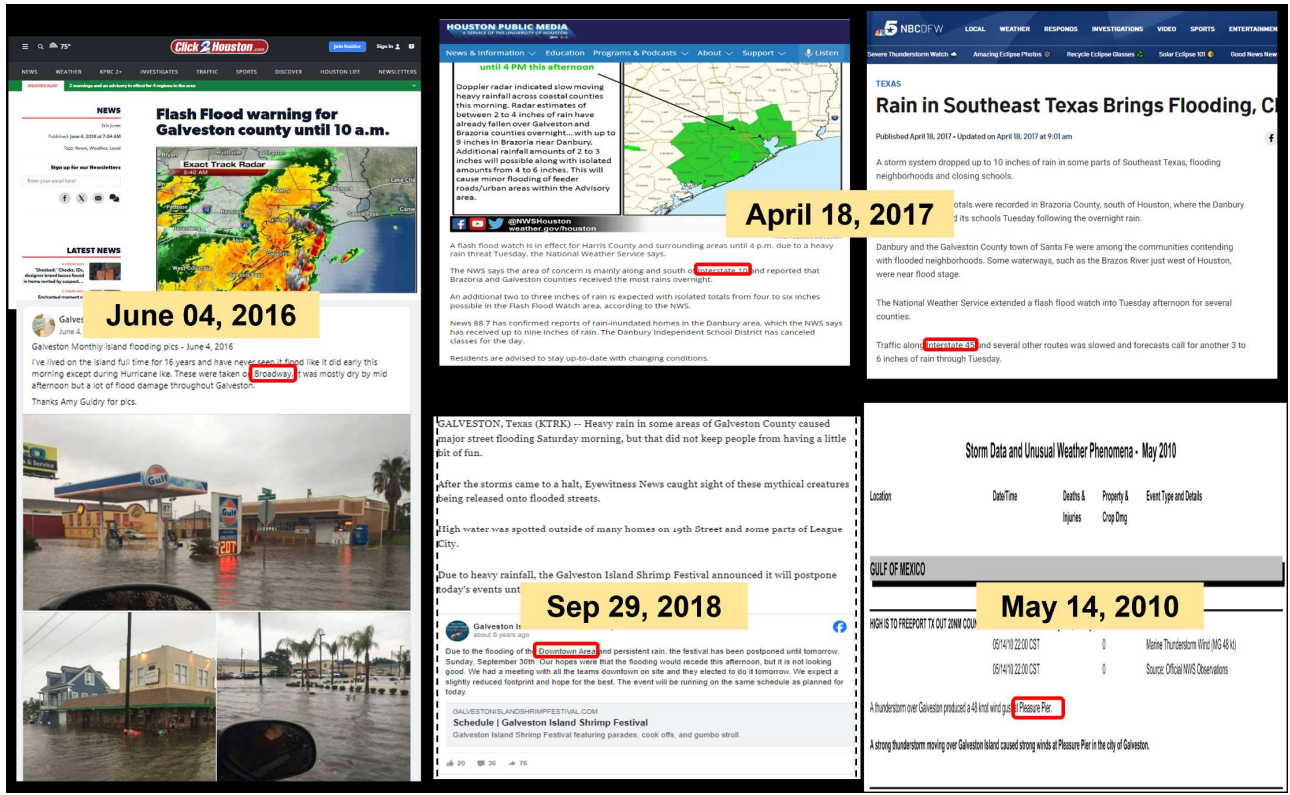

*Figure 6: Media snippets of high-water records in the flood events*

## 4.7 Rainfall Scenario Simulations

A unique capability of FlowsDT-Galveston is predictive modeling. Given the various intensities of rainfall events that occur, it is imperative to keep an updated purview of the effects of these rainfall events. This study showcases the effects of rainfall events of 2-year, 10-year, 25-year, 50-year and 100-year return period. The design rainfall scenarios were made by the symmetric method using Intensity-Duration-Frequency (IDF) curves with a duration of 24 hours. For representation, an additional comparative analysis is done amongst different return periods to assess maximum and mean flood levels in critical and non-critical structures. The IDF curves were obtained from NOAA's NWS Hydrometeorological Design Studies Center using the power-law model (Eqn. 5):

$$I = \frac{b}{(t_c + d)^e} \qquad (5)$$

where I is design rainfall intensity (in/hr), $t_c$ is the time of concentration in minutes, and e, b, d are the IDF coefficients for specific return frequencies. Table 1 presents the IDF coefficients used in this experiment.

*Table 1: Rainfall IDF coefficients[5] for Galveston County, TX*

| Return Period | b | d | e |
|---|---|---|---|
| **2-year** | 58.3 | 11.04 | 0.7839 |
| **10-year** | 70.47 | 12.06 | 0.7636 |
| **25-year** | 91.45 | 14.79 | 0.743 |
| **50-year** | 99.26 | 14.85 | 0.7308 |
| **100-year** | 115.89 | 16.5 | 0.7295 |

## 5   Results and Discussion

FlowsDT-Galveston has several uses and implications with regards to incorporating different fields of studies utilizing this model, promoting accelerated discovery. Its primary utilities are three fold: (1) FlowsDT-Galveston can be inputted with design rainfall scenarios using Soil Conservation Service (SCS) or Symmetric method using IDF curves to obtain spatiotemporal flood risk (scenario analysis); (2) Structural interventions in the form of Pipe Sizing or Low Impact Developments (like bio-retention tanks, rain garden, green roof, infiltration trench, permeable pavement, rain barrel, vegetative swales) can be modeled to evaluate their effectiveness (scenario analysis); (3) with real-time NEXRAD radar data through Radar Acquisition Project (RAP) in PCSWMM Real-Time (James et al., 2010), FlowsDT-Galveston can forecast inundations in real-time, substantially bolstering decision making during flooding. This section describes the developed digital twin platform in detail and demonstrates its utilities, including flood event modeling and validation, designed rainfall scenario simulation, and real-time forecasting.

---

[5] https://www.weather.gov/owp/hdsc

## 5.1 Immersive Visualizations in FlowsDT-Galveston

The final model along with its results were exported into StormCity for a 4D exploratory data analysis. Figure 7(a) presents the 3D flood depth and extent relative to building heights at a specific timepoint through realistic water flow textures. Furthermore, the velocity tracers compute the instantaneous calculated velocity and display it for the timestep chosen. This allows for a detailed examination of the dynamic interactions between floodwaters and urban structures. The interactive interface can show the results of the storm sewer system separately. Figure 7(b) offers an underground view of the storm sewer system, highlighting the flow and depth within conduits and junctions. An interactive popup is also displayed which allows for the selection of a specific conduit to graphically display its flow changes over time. It shows the various hydrologic and hydraulic properties (flow vs time, length of flow, etc.) of the highlighted conduit.

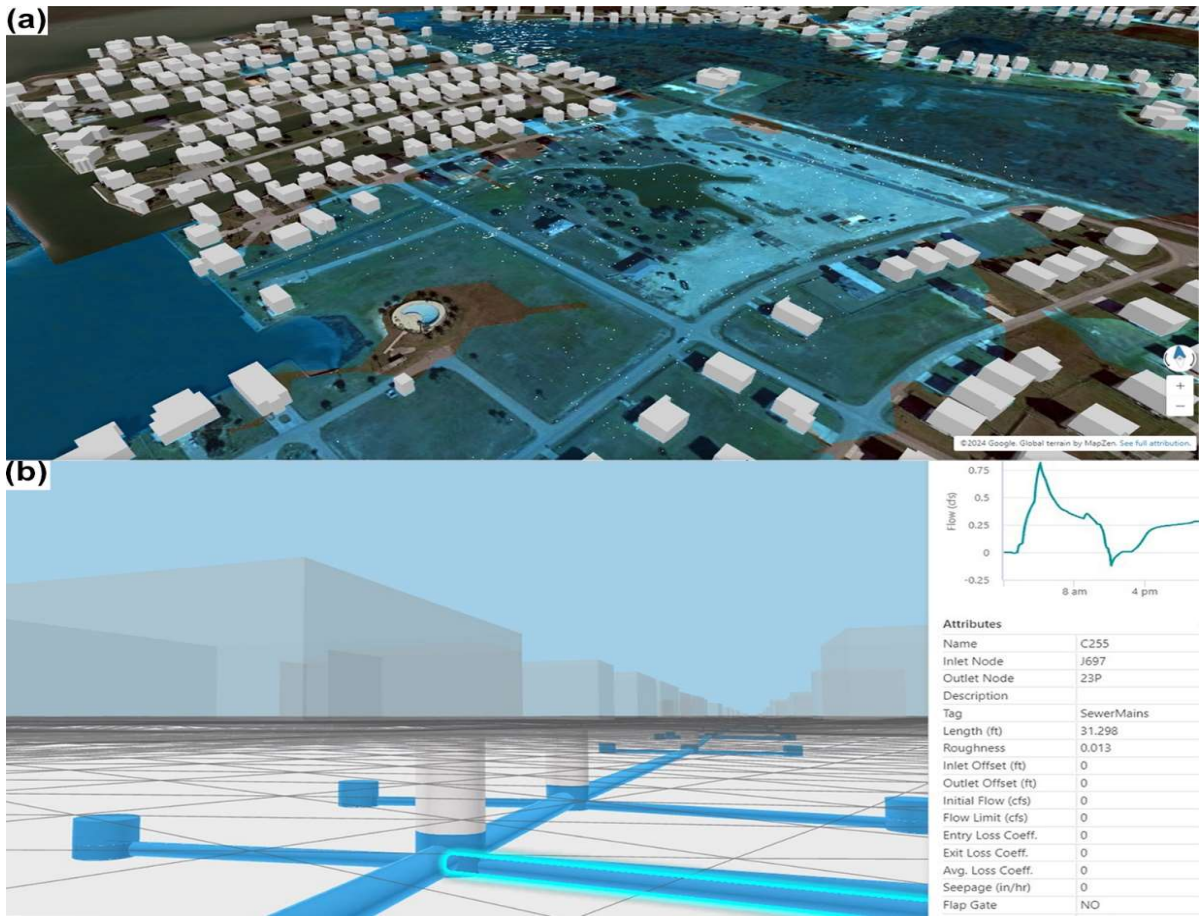

*Figure 7: Visualization of flood depths and flow in 3D environment*

In figure 8, a 4D visualization of flood depth and extent is displayed. Each time step is annotated separately at the top and has a time slider at the bottom of each instance. A video of the happenstance is also embedded for perusal. The 4D visualization helps in risk perception amongst the stakeholders by providing a multi-dimensional perspective of potential flood scenarios, enabling a more comprehensive understanding of their impacts on urban and rural landscapes. In other words, it facilitates more effective communication and collaboration among emergency responders, urban planners, government officials, and the public. By presenting information in a format that is intuitively understandable, it ensures that all parties can visualize and grasp the potential outcomes and necessary responses without ambiguity.

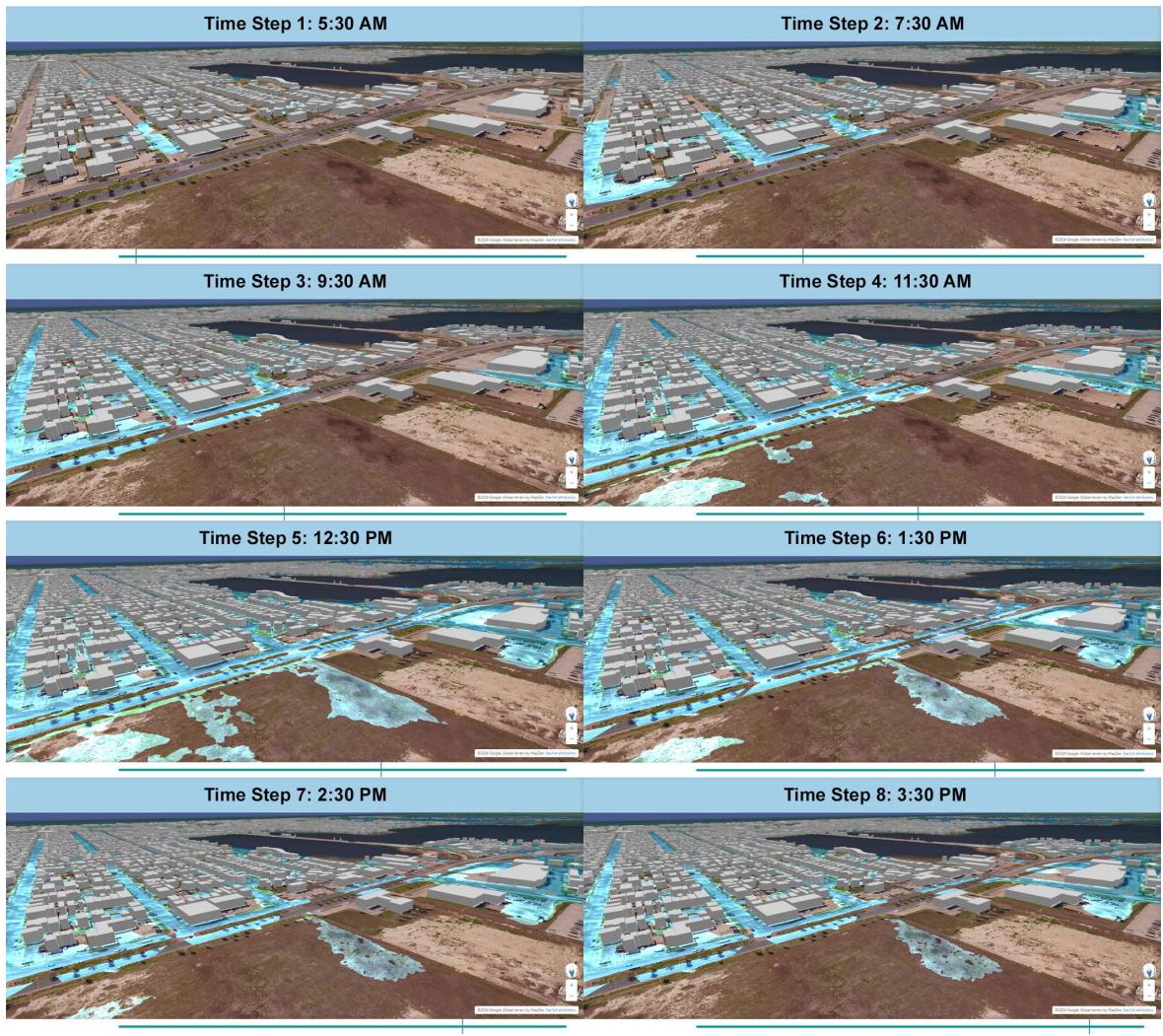

*Figure 8: Visualization of flood in the 4D environment of StormCity (Video attached)*

## 5.2 Historical Flood Events Simulation

The developed digital twin can effectively simulate historical flood events. The results from the hydrologic and hydraulic models were correlated with the elevation of Galveston to obtain spatiotemporal spread of the flood depths. Figure 9 displays the results of maximum flood depths of the four flood events at Northern Galveston. It is important to note that all flood models inherently carry a certain degree of uncertainty regarding flood depth, velocity and extent results. The uncertainty can be quantified by validating it with real-time flood events in future studies. The buildings are displayed in grey while the flood depths are classified by 1 ft. The results show that for all the events – the region in and around Harborside Drive and the southwest corner of the Offatts Bayou, were continuously getting flooded. Among these the least intense flooding happened on 18/04/2017 and the most intense on 05/14/2010 with respect to Northern Galveston region.

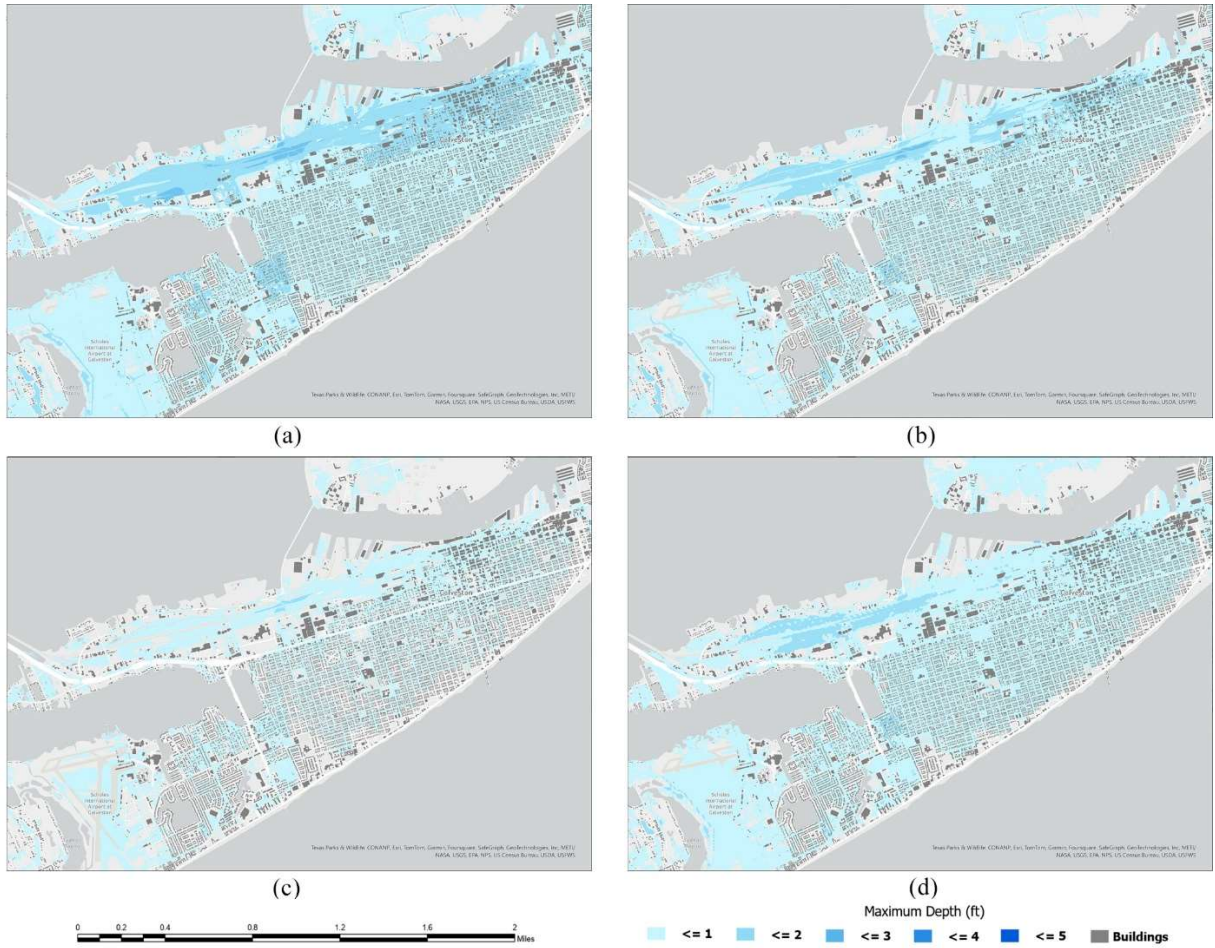

*Figure 9: Maximum Flood Depth results at Northern Galveston at (a) 05/14/2010, (b) 06/04/2016, (c) 18/04/2017, and (d) 09/29/2018*

Figure 10 displays the spatial patterns of simulated maximum flood velocity during the four flood event scenarios. The classification was based on geometric intervals. While the majority of the higher flood velocities were along the streets or outfalls, this does not seem to be a major issue as Galveston has a predominantly flat topography, and the range of velocities is low. Broadway and Harborside Drive have the highest velocities inland apart from regions where the flood water flowing out into the ocean.

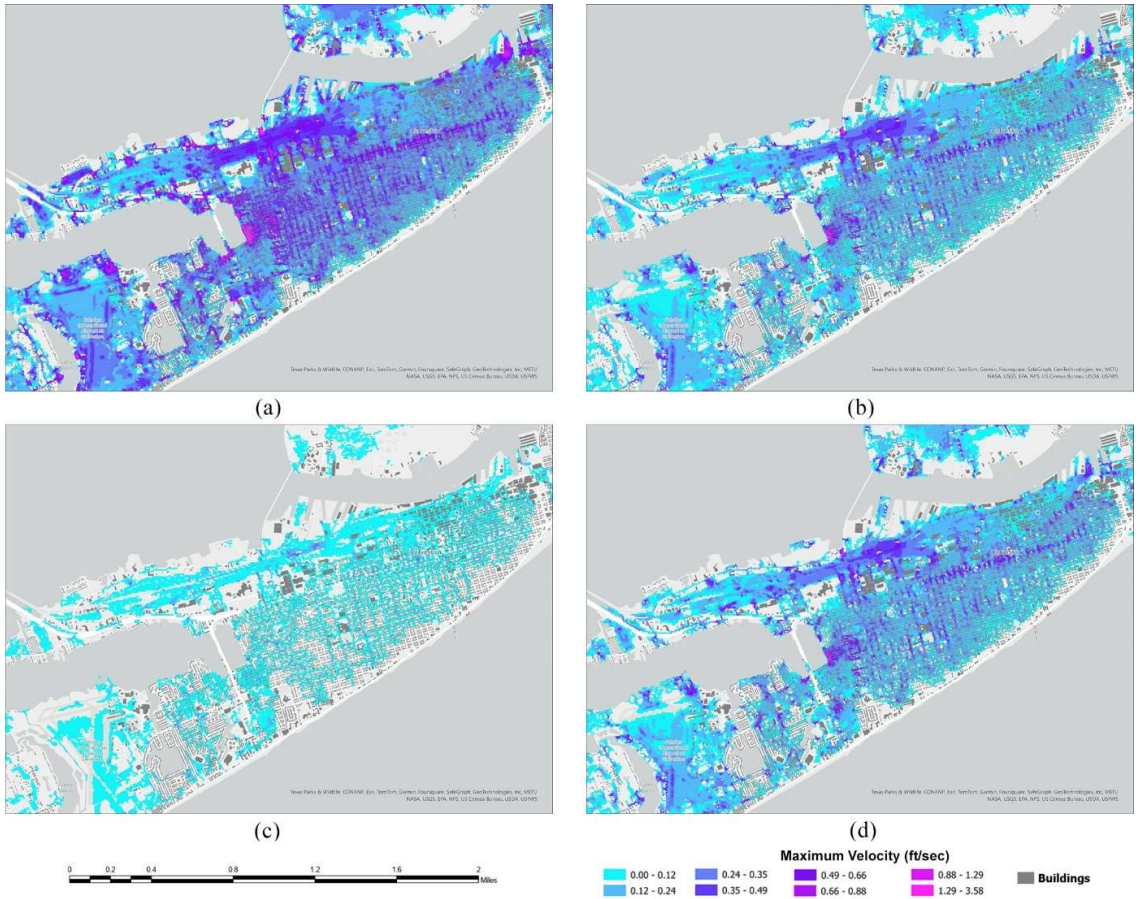

*Figure 10: Flood Velocity results at Northern Galveston at (a) 05/14/2010, (b) 06/04/2016, (c) 18/04/2017, and (d) 09/29/2018*

For brevity, the complete maximum depth result of one flood event (06/04/2016) is showcased in figure 11a.11 Figure 11b depicts the temporal stamps of the model's predictions on a dense section of Galveston City highlighting the speed at which various urban areas might be affected, offering windows for response and mitigation efforts. A video of this spread is linked to this figure.

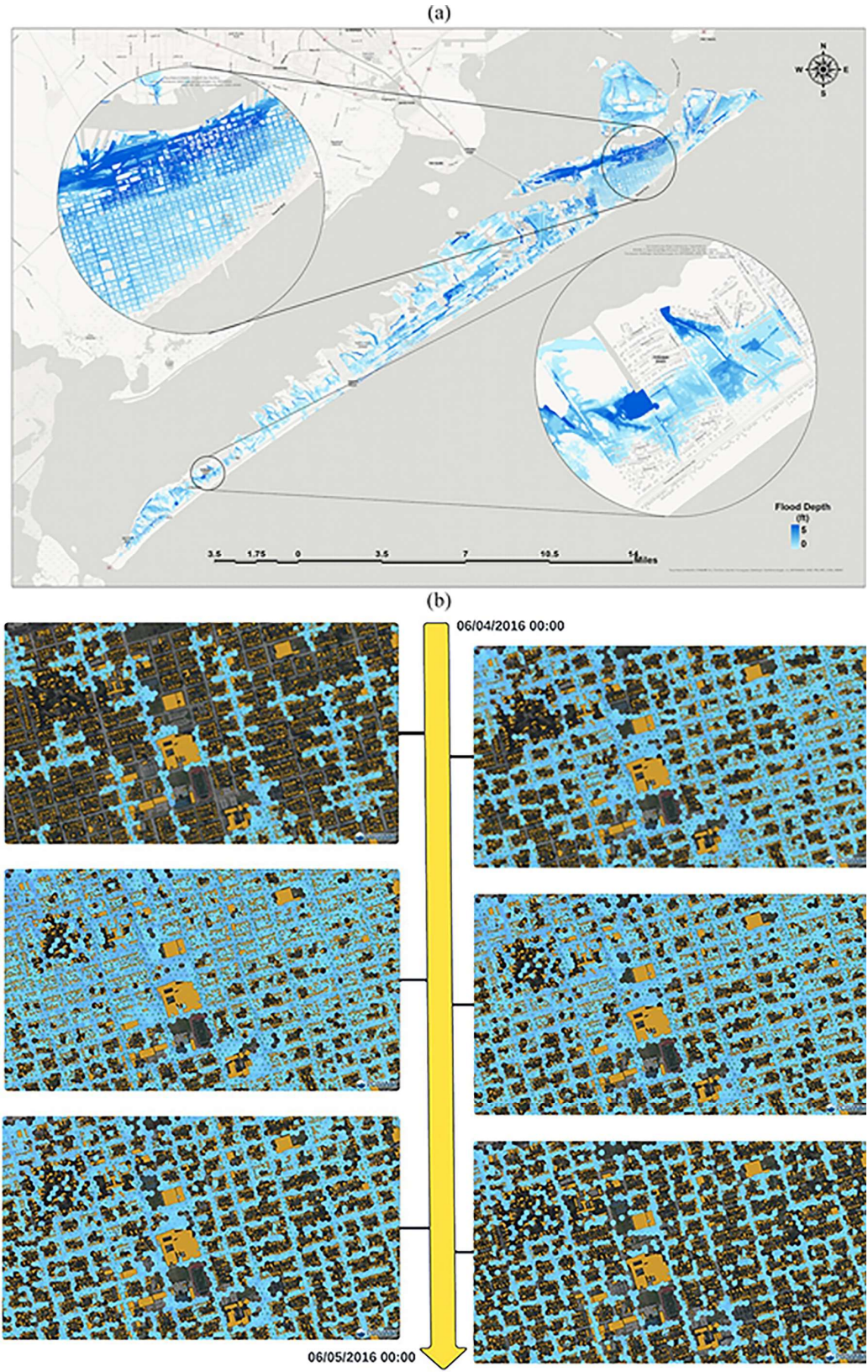

Figure 12: Top: Simulated maximum flood depth during the June 4, 2016, flood in Galveston, Texas, USA, Bottom: Temporal Pattern of flood depths in a densely habituated zone of Galveston City. (Video attached)

## 5.3    FlowsDT-Galveston Model Validation

Figure 12 contains the summary statistics of the validation data obtained by the integrated information harvesting system. The map displays the spatial distribution of the identified locations. For instance, on May 14, 2010, high water levels were documented on Stewart Road. On June 4, 2016, the flooding incident affected multiple areas including Harborside Drive, the lower portion of Greens Bayou, and Broadway. High flood locations were more extensively recorded on April 18, 2017, impacting the Galveston Island West End, Galveston Causeway, Galveston Pier 21, Pelican Island, Scholes Field, Offatts Bayou, Galveston Pleasure Pier, Moody Gardens, and Interstate 45. For September 29, 2018, high water occurrences were recorded in the heart of the city, specifically Downtown Galveston, the Galveston Strand, Broadway, and Market Street. Incidentally, as the majority of the northern part of Galveston, all of the retrieved locations were from this area, as denoted in Figure 12. We conducted a comprehensive evaluation across all specified locations and observed that each exhibited a considerable depth of flooding during the modeled events (A maximum of 3.6 ft for flooded nodes).

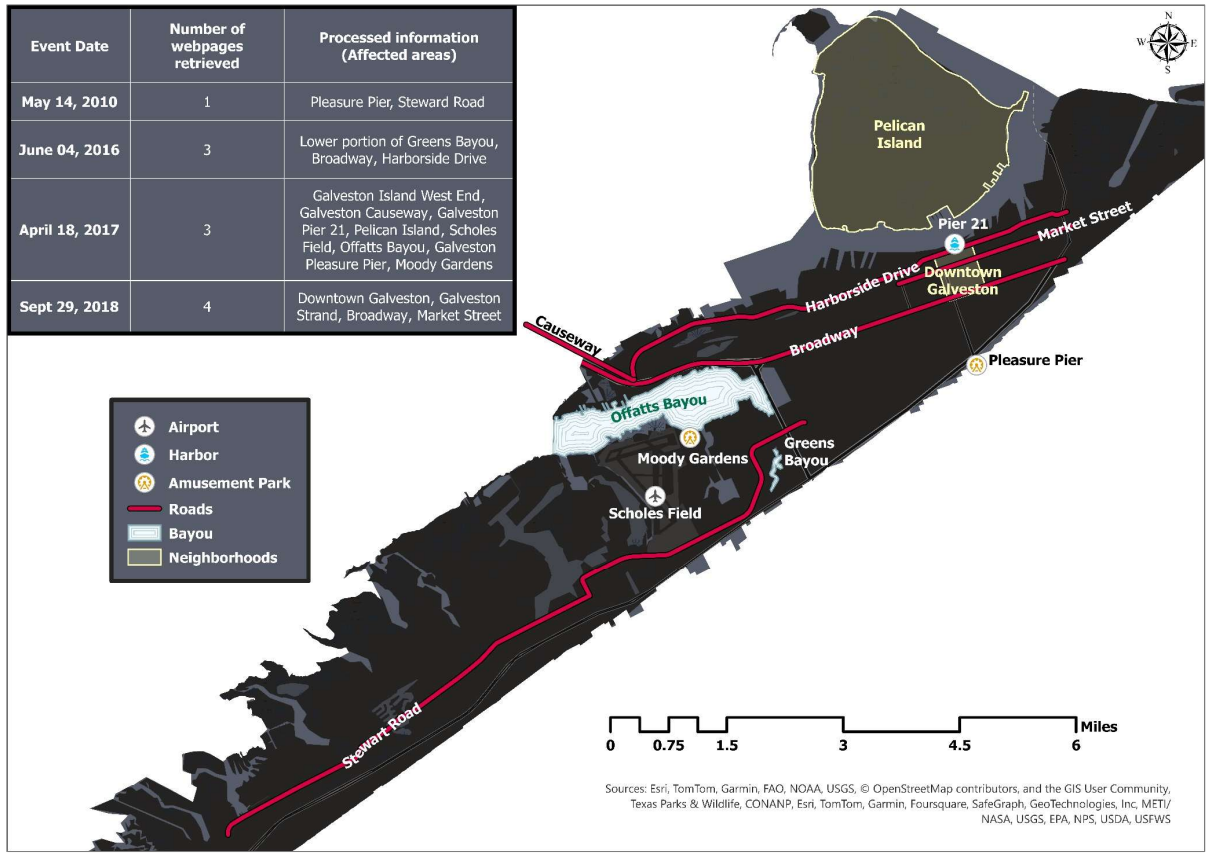

*Figure 132: Validation Statistics and Spatial Distribution*

## 5.4  Design Rainfall Scenario Analysis

To better understand the pattern of flood depths under various rainfall scenarios, inundated areas were categorized into five depth classes (<=1 ft, >1-2 ft, >2-3 ft, >3-4 ft, and >4-5 ft). As expected, the 100-year return period scenario produced the most widespread flooding, with over 720 million ft² affected, primarily in the shallow depth category (<=1 ft). As the return period decreased, both the extent and severity of inundation declined sharply. For example, areas inundated with more than 2 feet of water dropped from approximately 2.3 million ft² under the 100-year event to under 7k ft² in the 2-year scenario. Notably, the 10-year and 25-year events produced nearly identical inundation patterns across all depth classes, suggesting a threshold effect in the drainage system's capacity. These patterns show the city's vulnerability to extreme events and its relatively higher resilience to more frequent but less intense storms. Detailed values are provided in Table 2.

*Table 2: Flooded Area in different rainfall scenarios, not including building areas*

| Flooded Area (sq.ft) (*Not including obstructions*) | 100-year 24-hr | 50-year 24-hr | 25-year 24-hr | 10-year 24-hr | 2-year 24-hr |
|---|---|---|---|---|---|
| <=1 ft flood | 721,405,134 | 65,118,639 | 60,678,165 | 60,678,165 | 53,094,325 |
| >1 to <=2 ft flood | 61,929,071 | 4,823,915 | 2,995,778 | 2,995,778 | 1,142,349 |
| >2 to <=3 ft flood | 2,296,695 | 145,791 | 36,454 | 36,454 | 7,048 |
| >3 to <=4 ft flood | 47,781 | 3,139 | 2,097 | 2,097 | 345 |
| >4 to <=5 ft flood | 1,216 | 73 | 12 | 12 | 0 |

Figure 13 spatially depicts the maximum flood depth for design rainfall scenarios in the most densely populated area of Galveston City. These scenarios studied together identify the areas more susceptible to flooding, such as the Harborside Drive area near the Port and the bottom-right corner of Offatts Bayou.

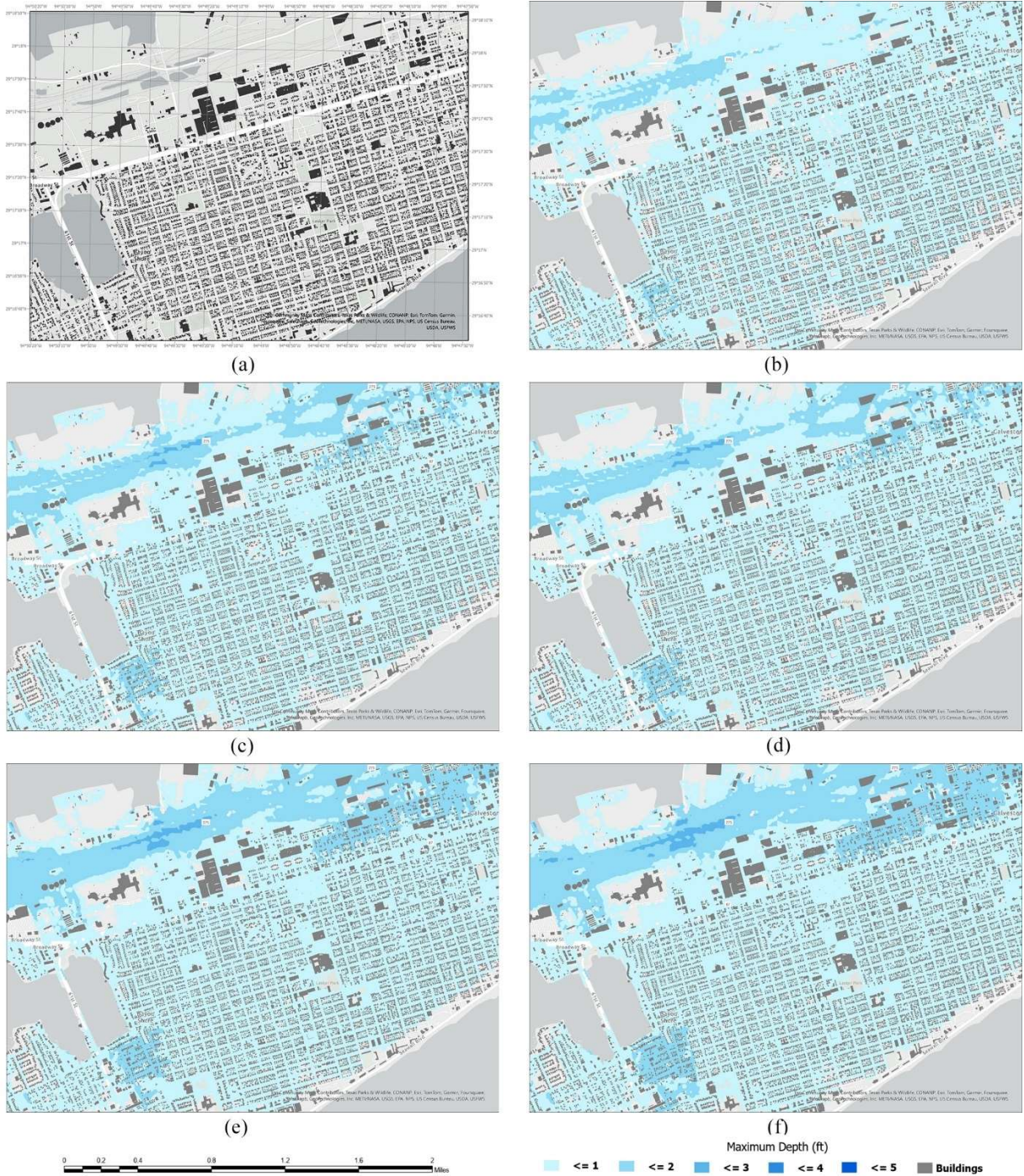

*Figure 14: Maximum flood depth (ft) for design rainfall scenarios – (a)Dry, (b)2Year-24hr, (c)10Year-24hr, (d)25Year-24hr, (e)50Year-24hr, and (f)100Year-24hr.*

The model therefore promotes locally informed and updatable comparative analysis at the building scale for the rainfall scenarios. For instance, using the LiDAR extracted building footprints, it is observed that the buildings that encountered the water flow. Table 3 showcases the numbers and areas of inundated buildings among a total of 28,244 buildings. Results show that buildings inundated by more than one foot increased from 0.5% in a 2-year flood to 6.2% in a 100-year flood over 24 hours. Overall, total buildings exposed to flooding depths up to 5 ft increase markedly from 19,146 (67.78%) in the 2-year scenario to 22,455 (79.5%) in the 100-year event.

*Table 3: Count and Area of Buildings impacted by flood inundation and their corresponding flood depths*

| Buildings | 100-year 24-hr | 50-year 24-hr | 25-year 24-hr | 10-year 24-hr | 2-year 24-hr |
|---|---|---|---|---|---|
| Buildings in <=1 ft flood | 20,715 (73.34%) | 20,712 (73.33%) | 20,177 (71.43%) | 20,177 (71.43%) | 19,003 (67.28%) |
| Buildings Area in <=1 ft flood (sq.ft) | 57,469,915 | 58,320,093 | 57,485,012 | 57,485,012 | 56,013,555 |
| Buildings in >1 and <=2 ft flood | 1,728 (6.11%) | 1206 (4.26%) | 677 (2.39%) | 677 (2.39%) | 141 (4.99%) |
| Buildings Area in >1 and <=2 ft flood (sq.ft) | 9,087,884 | 6,437,446 | 4,283,898 | 4,283,898 | 856,703.4 |
| Buildings in >2 and <=3 ft flood | 9 (0.03%) | 5 (0.01%) | 3 (0.01%) | 3 (0.01%) | 2 (0.00%) |
| Buildings Area in >2 and <=3 ft flood (sq.ft) | 56,466.65 | 171,971.8 | 232,474.3 | 232,474.3 | 222,235.1 |
| Buildings in >3 and <=4 ft flood | 2 (0.00%) | 1 (0.00%) | 1 (0.00%) | 1 (0.00%) | 0 (0%) |
| Buildings Area in >3 and <=4 ft flood (sq.ft) | 166,527.7 | 21,808.95 | 21,419.97 | 21,419.97 | 0 |

| Buildings in >4 and <=5 ft flood | 1 (0.00%) | 1 (0.00%) | 0 (0%) | 0 (0%) | 0 (0%) |
|---|---|---|---|---|---|
| Buildings Area in >4 and <=5 ft flood (sq.ft) | 21,419.97 | 21,419.97 | 0 | 0 | 0 |
| Total Buildings in <= 5 ft flood | 22,455 (79.5%) | 21,925 (77.62%) | 20,858 (73.84%) | 20,858 (73.84%) | 19,146 (67.78%) |
| Total Area in <= 5 ft flood (sq.ft) | 66,802,213 | 64,972,740 | 62,022,804 | 62,022,804 | 57,092,494 |

The hydrological impact for specific types of buildings can also be corroborated in this model. The Galveston City's GIS website[6] maintains a continuously updated database of property and building information. This along with our model delivers unique insights on infrastructural impacts of pluvial and fluvial flooding corresponding to – residential, economic, public utilities, bringing an edge to decision making among the residents, businessmen and city authorities alike.

Figure 14 showcases the spatial differences of velocity in the northern part of Galveston. For these scenarios it is mostly high in areas where the flood water drains to the ocean or along the major roads. As Galveston is a relatively flat city, the velocity of the water is seen to remain relatively stable in most parts. The highest flood velocity for a 100-year, 50-year, 25-year, 10-year, 2-year rainfall scenario was 3.59 ft/s, 3.13 ft/s, 2.69 ft/s, 2.32 ft/s, 1.92 ft/s respectively.

---

[6] https://gis-galveston.hub.arcgis.com/

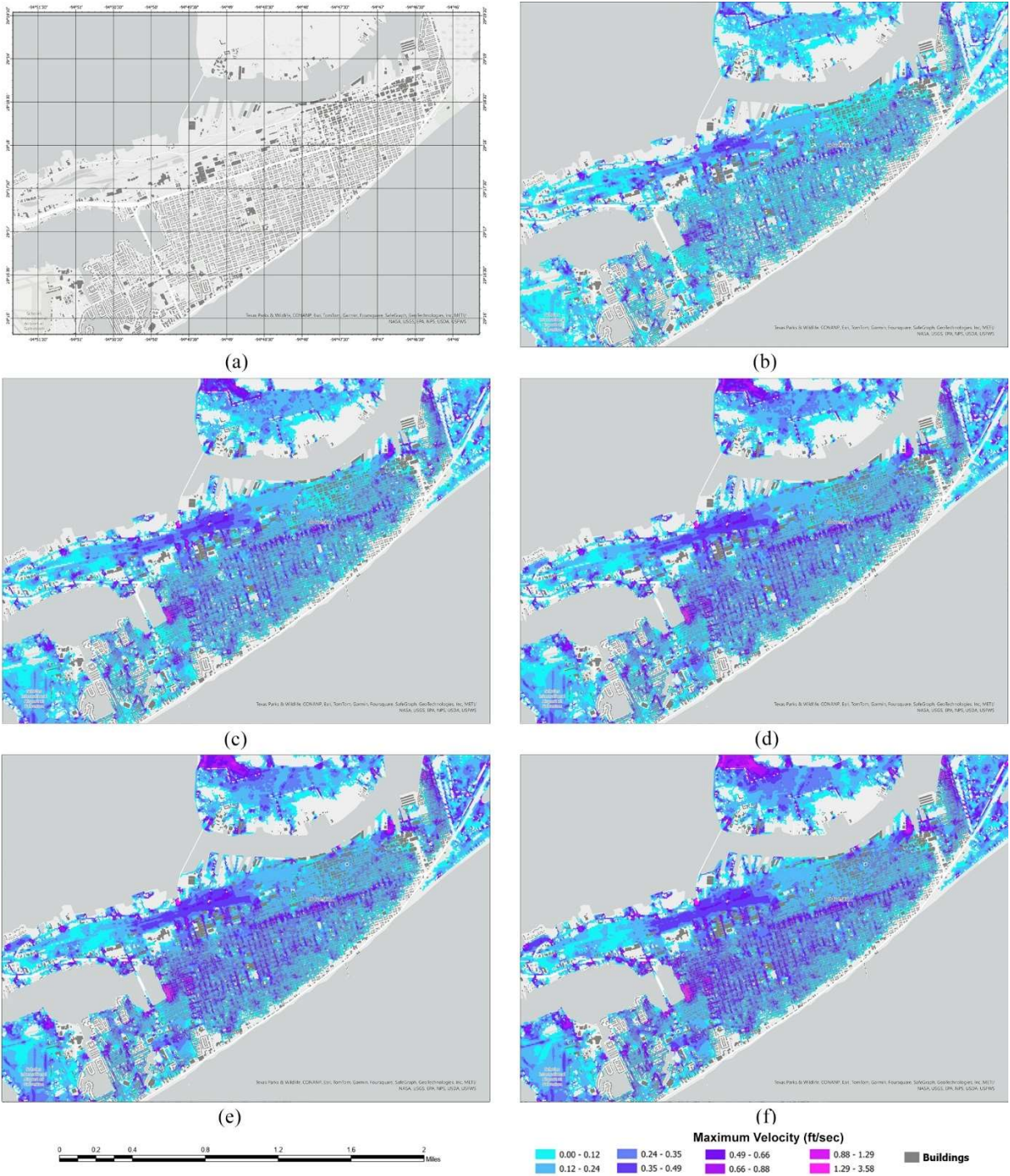

*Figure 154: Maximum flood velocity (ft/s) for design rainfall scenarios – (a)Dry, (b)2Year-24hr, (c)10Year-24hr, (d)25Year-24hr, (e)50Year-24hr, and (f)100Year-24hr.*

## 5.5    Real-Time Flood Forecasting

Real-time flood forecasting is a critical component of FlowsDT-Galveston. This component enables early warning systems to function more effectively, by allowing for more timely alerts to communities and emergency management teams. This advance notice is crucial for implementing evacuation plans, safeguarding critical infrastructure, and mobilizing resources efficiently. It supports decision-makers in making informed choices during critical periods, allowing them to prioritize interventions and allocate resources more strategically.

An advanced real-time flood forecasting system was implemented in FlowsGT-Galveston by integrating the RAP (Radar Acquisition Project). This integration facilitates real-time forecast of the extent, depth, and velocity of pluvial and fluvial flooding events. For data acquisition, a connection was established with the NEXRAD radar located in Galveston, Texas. As illustrated in Figure 15, this setup extracts precipitation information from temporally high-resolution radar data in CSV format. The hydrologic and hydraulic model uses precipitation to simulate event durations and provide high-resolution flood forecasts in real-time. Whilst the real-time data feed from the NEXRAD system allows for continuous monitoring and analysis of meteorological conditions, discrepancies might arise due to radar beam attenuation, reflection, or other atmospheric anomalies. To overcome this issue and ensure accuracy and reliability, the data was ground-truthed with in-situ rain gauge measurements (Scholes rain gauge). Therefore, by incorporating atmospheric data inputs directly into FlowsGT-Galveston, it was ensured that the flood forecasting models are both dynamic and robust, reflecting the latest changes in weather patterns and hydrological conditions.

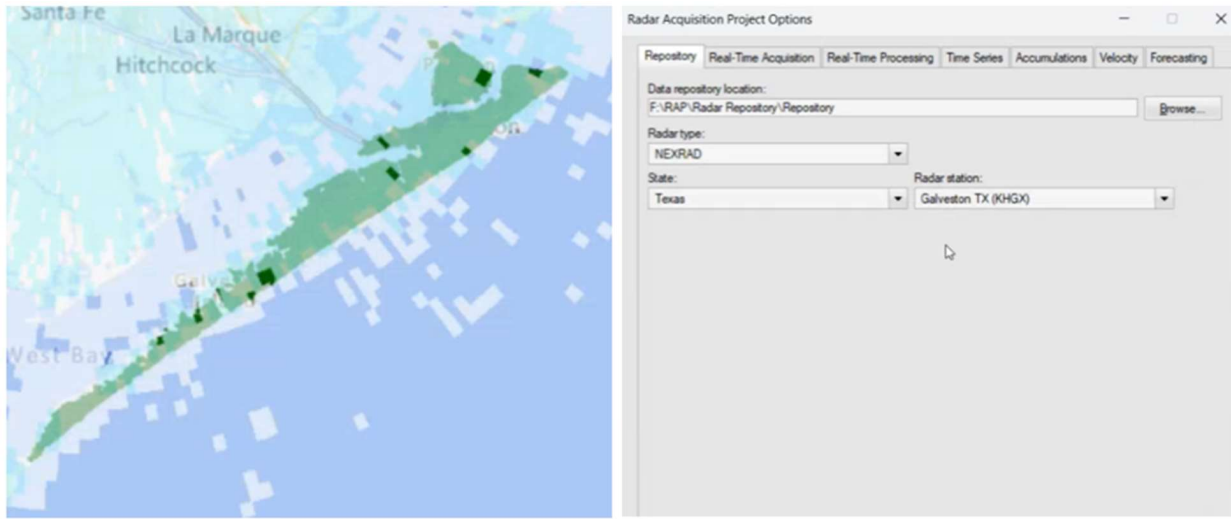

*Figure 165: Radar Data Acquisition of Galveston NEXRAD radar.*

## 6    Conclusion

This study presents the development and application of FlowsDT-Galveston, a high-resolution geospatial digital twin designed to model, monitor, and manage urban flood risks across the city of Galveston. FlowsDT-Galveston developed in this study is a platform combining high-resolution hydrologic-hydraulic simulations with real-time weather information, infrastructure analytics, and 4D visualization for urban flood modeling, management, and resilience improvement. By integrating 1D-2D coupled hydrodynamic simulations with LiDAR-derived topography, detailed infrastructure data, and real-time precipitation inputs combined using StormCity-based 4D interface, provides sub-meter resolution flood forecasts, enabling both historical reconstructions and predictive scenario analyses. This integration enables both analysis of historical flood reconstruction and predictive scenario for varied rainfall intensity testing, offering insights for disaster preparedness, infrastructure design, and climate adaptation planning.

In this study, FlowsDT-Galveston is tested with four significant historical flood events in Galveston, TX (on May 14, 2010; June 4, 2016; April 18, 2017; and September 29, 2018). Simulated outcomes for the aforementioned events shows that the region such as Harborside Drive, Broadway Avenue, and Offatts Bayou are continuously getting inundated leading to one of the few flood hotspots across Galveston. These events locations are additionally cross validated with crowdsource information for the heavy rainfall events. Model validation against high-water marks and geolocated flood reports demonstrated strong correlation, particularly in the densely urbanized corridors of the city. An integral part of FlowsDT-Galveston, StormCity-based 4D interface, facilitated hourly visualizations of flood inundation, enhancing understanding of flood dynamics and creating an integrated tool for stakeholders, thereby enabling near-real-time decision support. In order to study the robustness of the structures and identify the potential areas subjected to flood with different rainfall patterns, FlowsDT-Galveston is further analyzed for different return periods. FlowsDT-Galveston produced high-resolution flood inundation maps across multiple return periods: 2-, 10-, 25-, 50-, and 100-year for a storm duration of 24 hours. With recent changes in patterns of rainfall intensities and changing climate, 100-year return period is used to analyze the effectiveness of the model (Martel et al., 2021). Under a 100-year storm scenario, approximately 721 million ft² of Galveston land area was inundated with water depths up to 1 foot, while 2.3 million ft² experienced depths exceeding 2 feet. The number of buildings affected by flooding increased from 19,146 (67.78%) in the 2-year event to 22,455 (79.5%) in the 100-year event. Buildings experiencing flood depths greater than 1ft rose from 0.5% to 6.2% over this range, indicating an escalation of flood risk with rainfall. In addition to depth-based impacts, flood velocity analysis revealed concentrated flow paths along major urban roads and coastal outfalls.

The FlowsDT-Galveston study provides valuable insights into urban flood dynamics, but several limitations highlight opportunities for future research. The model's fidelity is highly dependent on the quality and resolution of input datasets, particularly land use, topography, and stormwater infrastructure, underscoring the need for continuous data collection, sensor integration, and routine model updates to improve accuracy. Future research should incorporate building classifications, such as residential, commercial, and critical infrastructure (e.g., schools, hospitals, power systems), to better assess the cascading societal and economic effects of floods. Finally, extending FlowsDT-Galveston to evaluate the effectiveness of various intervention measures presents a promising direction for future work, supporting informed decision-making for flood resilience planning. While the developed Geospatial DT offers robust tools for urban flood resilience, it is important to acknowledge its computational burdens. In the current configuration of our setup (Intel i9 Processor @ 3.70GHz, NVIDIA GeForce RTX 3090, 64 GB RAM), the model had taken 13:24 hours to simulate the 100-year 24-hour rainfall. The current model's effectiveness is contingent on the availability and integration of up-to-date data. As urban landscapes and climatic conditions evolve, so must the data that feeds the Geospatial DT. Continuous improvement and regular updates are essential to maintain or improve the accuracy and relevance of the model. This requires ongoing surveys and data collection through Internet of Things (IoTs) and determining an optimum number of sensors required based on the spatial variation of the required data. Moreover, the Geospatial DT's current framework may not fully capture the rapid socio-economic changes and their impacts on urban infrastructure, necessitating further additions to convert this from a Digital Twin Aggregate to a Digital Twin Environment. Summarily, FlowsDT-Galveston serves as a modular and extensible platform vetted against historical data for urban flood modeling and resilience planning. By integrating different scenario simulations, real-time forecasting capabilities, return period based flood analysis, crowdsourced validation, and effective visualizations, it offers a powerful decision-support tool that enhances the traditional flood model capabilities. This study therefore illustrates the potential of digital twin technology in informing urban resilience strategies and the architecture outlined in this study can serve as a transferable blueprint for cities worldwide experiencing hydroclimatic extremes and urbanization pressures.

## Funding Acknowledgements

This research is based on work supported by the Computational Hydraulics Int. (CHI) Educational Grant, National Academies of Sciences, Engineering, and Medicine (NASEM) under the Gulf Research Program (SCON-10000653), and National Science Foundation (2122054, 2318206).